\documentclass[prx,floatfix,twocolumn,tightenlines,nofootinbib,superscriptaddress]{revtex4-1}

\usepackage[utf8]{inputenc}
\usepackage[T1]{fontenc}
\usepackage{graphicx}
\usepackage{amsmath,amssymb,amsthm,amsfonts}
\usepackage{psfrag}
\usepackage{ifsym}
\usepackage{dsfont}
\usepackage{enumerate}
\usepackage{multirow}
\usepackage{hhline}
\usepackage{bbm}
\usepackage[colorlinks=true,linkcolor=red,citecolor=blue]{hyperref}
\usepackage{float}
\usepackage[sort&compress]{natbib}
\usepackage[dvipsnames]{xcolor}
\usepackage{soul}
\usepackage{comment}
\newcommand{\myop}[1]{\ensuremath{\operatorname{#1}}}

\newcommand{\ket}[1] {| #1 \rangle}

\newcommand{\expec}[1]{\left\langle #1 \right\rangle}
\newcommand{\one}{\leavevmode\hbox{\small1\normalsize\kern-.33em1}}

\newcommand{\id}{\mathds{1}}

\DeclareMathOperator{\tr}{tr}

\begin{document}
\title{Measurement Device Independent Verification of Quantum Channels}

\author{Francesco Graffitti}
\email[Corresponding author: ]{fraccalo@gmail.com}
\affiliation{Institute of Photonics and Quantum Sciences, School of Engineering and Physical Sciences, Heriot-Watt University, Edinburgh EH14 4AS, UK}

\author{Alexander Pickston}
\affiliation{Institute of Photonics and Quantum Sciences, School of Engineering and Physical Sciences, Heriot-Watt University, Edinburgh EH14 4AS, UK}

\author{Peter Barrow}
\affiliation{Institute of Photonics and Quantum Sciences, School of Engineering and Physical Sciences, Heriot-Watt University, Edinburgh EH14 4AS, UK}

\author{Massimiliano Proietti}
\affiliation{Institute of Photonics and Quantum Sciences, School of Engineering and Physical Sciences, Heriot-Watt University, Edinburgh EH14 4AS, UK}

\author{Dmytro Kundys}
\affiliation{Institute of Photonics and Quantum Sciences, School of Engineering and Physical Sciences, Heriot-Watt University, Edinburgh EH14 4AS, UK}

\author{Denis Rosset}
\affiliation{Perimeter Institute for Theoretical Physics, 31 Caroline Street North, Waterloo, Ontario, Canada N2L 2Y5}

\author{Martin Ringbauer}
\affiliation{Institut f\"{u}r Experimentalphysik, Universit\"{a}t Innsbruck, Technikerstrasse 25, A-6020 Innsbruck, Austria}

\author{Alessandro Fedrizzi}
\affiliation{Institute of Photonics and Quantum Sciences, School of Engineering and Physical Sciences, Heriot-Watt University, Edinburgh EH14 4AS, UK}

\begin{abstract}
The capability to reliably transmit and store quantum information is an essential building block for future quantum networks and processors. Gauging the ability of a communication link or quantum memory to preserve quantum correlations is therefore vital for their technological application. Here, we experimentally demonstrate a measurement-device-independent protocol for certifying that an unknown channel acts as an entanglement-preserving channel. Our results show that, even under realistic experimental conditions, including imperfect single-photon sources and the various kinds of noise---in the channel or in detection---where other verification means would fail or become inefficient, the present verification protocol is still capable of affirming the quantum behaviour in a faithful manner without requiring any trust on the measurement device.
\end{abstract}

\maketitle

% =========================================================
% Introduction
% =========================================================
The ability to transmit and store quantum states, and coherently manipulate the timing of photonic signals is a crucial requirement in quantum technologies~\cite{RevModPhys.79.135}. Quantum communication links in combination with quantum memories form the quantum channels that offer these capabilities. These quantum channels thus play a pivotal role in enabling full scale quantum networks~\cite{kimble2008quantum}, promising unconditionally secure communication and the prospect of distributed quantum computing.

With the development of such quantum channels, especially quantum memories, comes the challenge of certifying their capabilities~\cite{PhysRevA.88.042335,Pusey:15}. 
In particular, we seek the ability to discern a truly non-classical channel from a cheap knock-off, such as a channel that simply measures the input state and approximately re-prepares it at the output.
While the latter could preserve some information about the state, it cannot preserve the exact quantum state nor any previously established correlations, rendering it useless for quantum applications.
We denote channels of this sort as entanglement-breaking (EB), in contrast to true quantum memories or coherent quantum channels, which preserve entanglement at least to some extent.

Consider an unknown channel that is claimed to be non-classical, i.e.\ entanglement preserving. The most straightforward approach to obtain a complete characterization of the channel is a tomographic reconstruction of the channel's process matrix~\cite{PhysRevLett.86.4195}. In practice, however, this approach is too resource intensive for all but the smallest channels and further requires precise control and trust in all parts of the experiment. In most cases, such trust is undesirable or cannot be guaranteed at all. One way to overcome this is by using the correlations of entangled quantum systems, where a violation of a Bell inequality certifies the presence of entanglement even when the measurements are performed by untrusted black-box devices. Consequently, Bell-test-based protocols allow for the verification of quantum channels in a so-called \emph{device-independent} way, that is without requiring any trust in the experimental devices~\cite{PhysRevLett.106.250404}, even providing guarantees on the quality of the channel via self-testing~\cite{Bancal2018}. On the flip side, these approaches cannot capture all non-classical channels~\cite{Pal_2015} and are subject to challenging loopholes~\cite{RevModPhys.86.419,PhysRevA.46.3646,Rosset2017} that make them very fragile to losses and experimentally difficult to implement. Moreover, in practice we rarely face the situation where nothing can be trusted such that a fully device-independent approach is necessary. Instead, while we might face an untrusted measurement device, we typically have access to a trust-worthy state preparation device allowing us to generate well-defined quantum states of our choice: a scenario commonly referred to as measurement-device-independent (MDI).

MDI schemes were first proposed in the context of entanglement verification~\cite{PhysRevLett.110.060405} for spatially separated subsystems, within the framework of semi-quantum games~\cite{Buscemi2012}, with several advantages over Bell tests~\cite{Rosset2013}. These methods have since been extended to quantum steering~\cite{Cavalcanti2013a} and to the analysis of entanglement structure~\cite{Zhao2016} and its quantification~\cite{Shahandeh2017, Supic2017,Rosset2017b}, which has been demonstrated experimentally~\cite{PhysRevLett.112.140506, Nawareg2015,Verbanis2015}. Moving beyond the classic scenario of two spacelike separated parties, the MDI approach and semi-quantum games can be adapted to situations where one party wants to test another party's ability to maintain quantum coherence over time~\cite{PhysRevX.8.021033}, such as in a quantum memory.

Here we demonstrate experimentally, that MDI verification of quantum memories and more general quantum channels is a simple technique that is highly robust to experimental imperfections and viable with current technology. We study the performance and success probability of the method for channels suffering from depolarizing and dephasing noise, taking into account all experimental imperfections (such as imperfect state preparation when multiple copies of the input state are prepared), a problem that so far hasn't received sufficient attention. Under all conditions achievable with current technology, we find that the MDI approach outperforms Bell-test based techniques in terms of resource requirements as well as noise resilience without the need for extra assumptions such as fair sampling. This method can thus certify a much wider range of channels, and remains practical under realistic experimental conditions.

We now consider a typical experimental scenario, where a client (Alice) wishes to test a potentially dishonest quantum memory (provided by Bob) before deployment in a quantum network. Alice is assumed to possess a trusted preparation device, which is a scenario that naturally lends itself to the use of semi-quantum games. Here, Alice repeatedly asks Bob a set of randomized ``questions'' by sending him quantum states and gets back a classical answer from Bob in every round. Bob is then asked to maximize a payoff function chosen to witness whatever quantum property Bob claims to possess. Since the questions are non-orthogonal, Bob merely knows the set of possible questions, but cannot know which question is asked in each round and thus cannot cheat. This method thus allows Alice to witness whether Bob possesses the claimed quantum property without having to trust him.

In each round, Alice sends successively two questions with a time delay between them, which forces Bob to store the first question until the second one arrives. In our notation, the first question is a state chosen at random from a finite set $\{\xi_x\}$ indexed by $x$, while the second question is chosen from a finite set $\{\psi_y\}$ indexed by $y$; both questions are sampled with uniform probability. After receiving the second question, Bob returns a classical answer $b$ back to Alice. Bob is asked to maximize a pre-arranged payoff function $\omega(b,x,y)$ using the quantum channel $\mathcal{N}$ at his disposal. In analogy with Bell scenarios, we write the payoff then achieved $W = \sum_{bxy} \omega(b,x,y) P(b|xy)$. The combination of the coefficients $\omega(b,x,y)$ with the sets $\{\xi_x\}$ and $\{\psi_y\}$ is a temporal semi-quantum game~\cite{Rosset2017}. Every such game has an upper bound $W_\text{EB}$ on the payoff achievable when Bob has only an entanglement breaking channel at its disposal.

\begin{figure}[t!]
    \begin{center}
    \includegraphics[width=0.9\columnwidth]{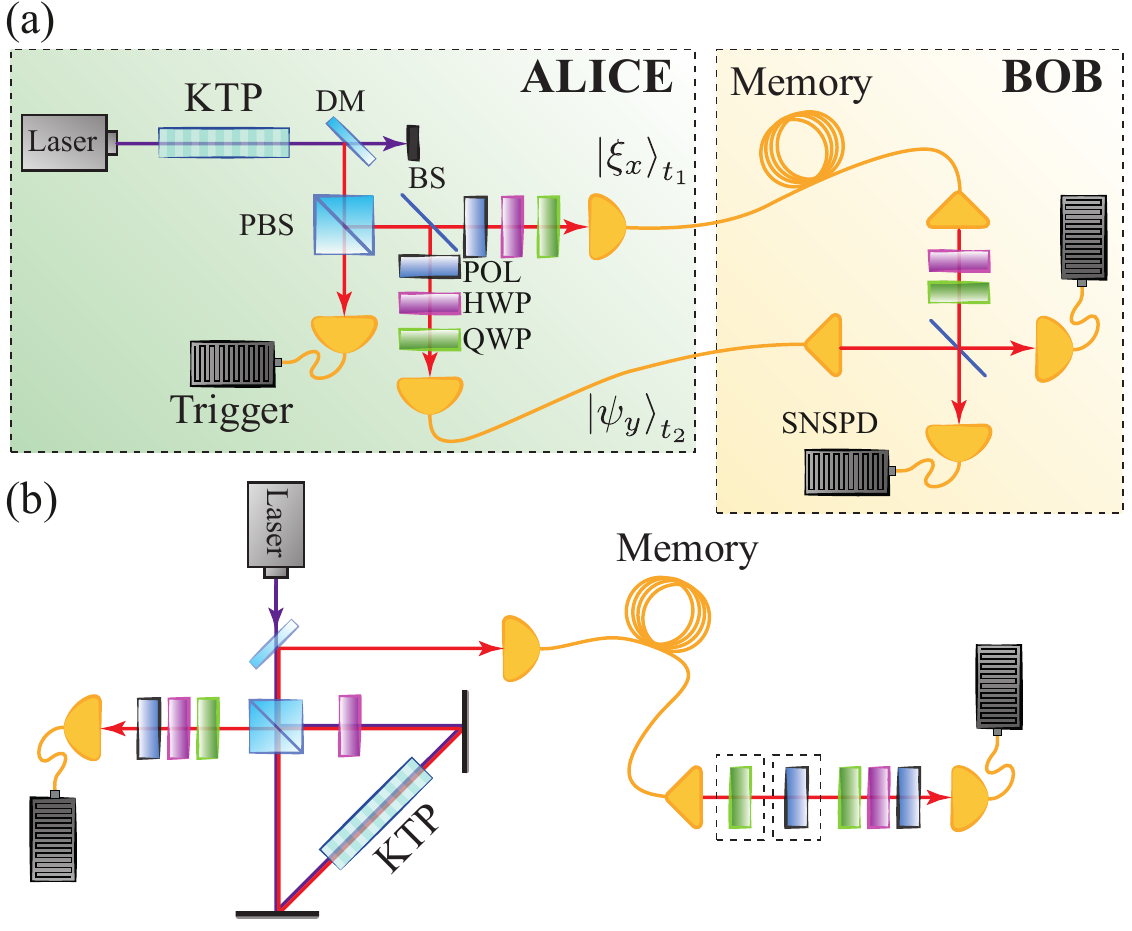}
    \end{center}
    \vspace{-2em}
    \caption{\textbf{Experimental Setup.}
    (a) MDI entanglement witness setup. 
    Pairs of 1550 nm single photons are generated via degenerate PDC in a custom-poled potassium titanyl phosphate (KTP) crystal~\cite{graffitti2017pure,graffitti2018independent} pumped by 775 nm, 1.6 ps pulses with 80 MHz repetition rate. 
    After being separated from the pump with a dichroic mirror (DM), one photon of each pair is transmitted on a polarizing beam-splitter (PBS) and detected by a superconducting nanowire single-photon detector (SNSPD), providing the heralding signal for its twin photon. The heralded photons are manipulated using a sequence of a polarizer (POL), half-wave (HWP), and quarter-wave plate (QWP) to prepare the questions of the MDI protocol. 
    The left-green (right-yellow) shading indicates the trusted (untrusted) parts of the experiment. 
    (b) Untrusted channel verification setup via Bell test.
    }
    \label{fig:Setup}
\end{figure}

In our experiment, the sets $\{\xi_x\}$ and $\{\psi_y\}$ are identical and composed of symmetric informationally-complete (SIC) single-qubit quantum states which form a non-orthogonal basis of the Hilbert space with constant pairwise overlap of $1/3$~\cite{Renes2004}. As shown in the Supplementary Materials, other set of states can be chosen for the protocol with some implementation benefits. Non-orthogonality ensure that, although Bob knows the set of possible questions, he cannot with certainty know which questions are being asked in each round of the game.
We write $x, y = 0,1,2,3$ the indices of these two successive questions $\xi_x$, $\psi_y$ sent to Bob, while Bob sends back a classical answer $b = 0,1$.
The property we are testing is a claim made by Bob that he possesses a non-entanglement-breaking channel corresponding to a Choi matrix $\Phi$~\cite{PhysRevA.87.022310}.
Accordingly, this Choi matrix is entangled, a fact that can be tested by an entanglement witness~\cite{GUHNE20091} $F$ such that $\tr{\left[F \Phi\right]} > 0$ while $\tr{\left[F \Phi_\text{EB}\right]} \le 0$ for Choi matrices of entanglement breaking channels.
Our payoff coefficients $\omega(b,x,y)$ are chosen so that $\omega(b=1,x,y) = 0$, while $\omega(b=0,x,y)$ provides a decomposition of that witness $F = \sum_{xy} \omega(0,x,y) \left(\xi_x^\top \otimes \psi_y^\top \right)$. This ensures~\cite{Rosset2017} that $W_\text{EB} = 0$ while $W > 0$ when Bob actually implements the channel he claims to possess and projects the joint two-photon state onto a singlet state $\ket{\Psi^{-}}$. In each round, Bob announces his result $b$ to Alice, who computes the payoff using her knowledge of the prepared questions. We studied the effects of an imperfect quantum channel by simulating additional depolarizing $\mathcal{N}_\text{P}$ noise or dephasing noise $\mathcal{N}_\phi$, defined as:
\begin{equation}
\begin{split}
\mathcal{N}_\text{P}(\rho) &= (1-p) \rho + p \id/2 ,\\
\mathcal{N}_\phi(\rho) &= (1-\frac{p}{2}) \rho + \frac{p}{2} \sigma_3 \rho \sigma_3 ,
\label{Eq:noise}
\end{split}
\end{equation}
where $\id$ and $\sigma_3$ are the identity and Pauli Z operator respectively, and $0 \le p \le 1$ is the noise strength. In both cases, the optimal payoff coefficients are found to be
\begin{equation}
  \begin{split}
    &\omega(b=0,x,y) = \begin{cases}
      -5/8 & \text{if } x=y\\
      1/8 & \text{otherwise}
    \end{cases}
    \\
    &\omega(b=1,x,y) = 0 ,
  \end{split}
  \label{eq:payof}
\end{equation}
where $b=0$ corresponds to a successful projection of the joint state onto the singlet state and $b=1$ to any other measurement outcome.

Experimentally, we use a heralded parametric down-conversion (PDC) source to emulate a single photon source at telecommunication wavelength (the experimental setup is shown in Fig.~\ref{fig:Setup} (a)). 
Alice encodes a probe state $\xi_x$ in the heralded photon using a sequence of POL, HWP and QWP. 
This probe state represents the first question in our semi-quantum game, which Bob receives at time $t_1$, and is asked to process in his alleged quantum channel. 
Bob's quantum channel is a 15 m single-mode fibre emulating a quantum memory with fixed storage time of $\sim$75 ns.
A HWP and a QWP are used to implement a noisy channel with variable noise-strength $p$ by applying a combination of Pauli operators according to Eq.~\eqref{Eq:noise} for a measurement time proportional to $p$.
At a later time $t_2$, Alice prepares in the same way a second probe state $\psi_y$---corresponding to the second question in the game---and sends it to Bob, who is asked to perform a joint measurement on the two states via two-photon interference on a beam-splitter (BS) and broadcast the outcome: only a coincidence click event of the detectors after the BS corresponds to $b=0$ in eq.~\eqref{eq:payof} (i.e. a successful projection on the singlet state), while any other event corresponds to $b=1$.

Unlike the MDI protocol described above, a Bell-test approach, i.e.\ a fully device-independent verification of a quantum channel, requires Alice to prepare entangled quantum state. We produce entangled pairs of photons via PDC in Sagnac interferometric scheme~\cite{Fedrizzi2007Sagnac}, as shown in Fig.~\ref{fig:Setup} (b).
Alice then sends one photon of the entangled pair to Bob, who sends it through his channel. An additional set of HWP, QWP, and POL is used to introduce controllable dephasing and depolarizing noise. After the stored photon has been retrieved a Clauser-Horne-Shimony-Holt (CHSH) Bell-test~\cite{PhysRevLett.23.880} is performed on the joint system, and a violation of the inequality guarantees the genuine quantumness of the channel.

\begin{figure}[h!]
  \begin{center}
    \includegraphics[width=0.95\columnwidth]{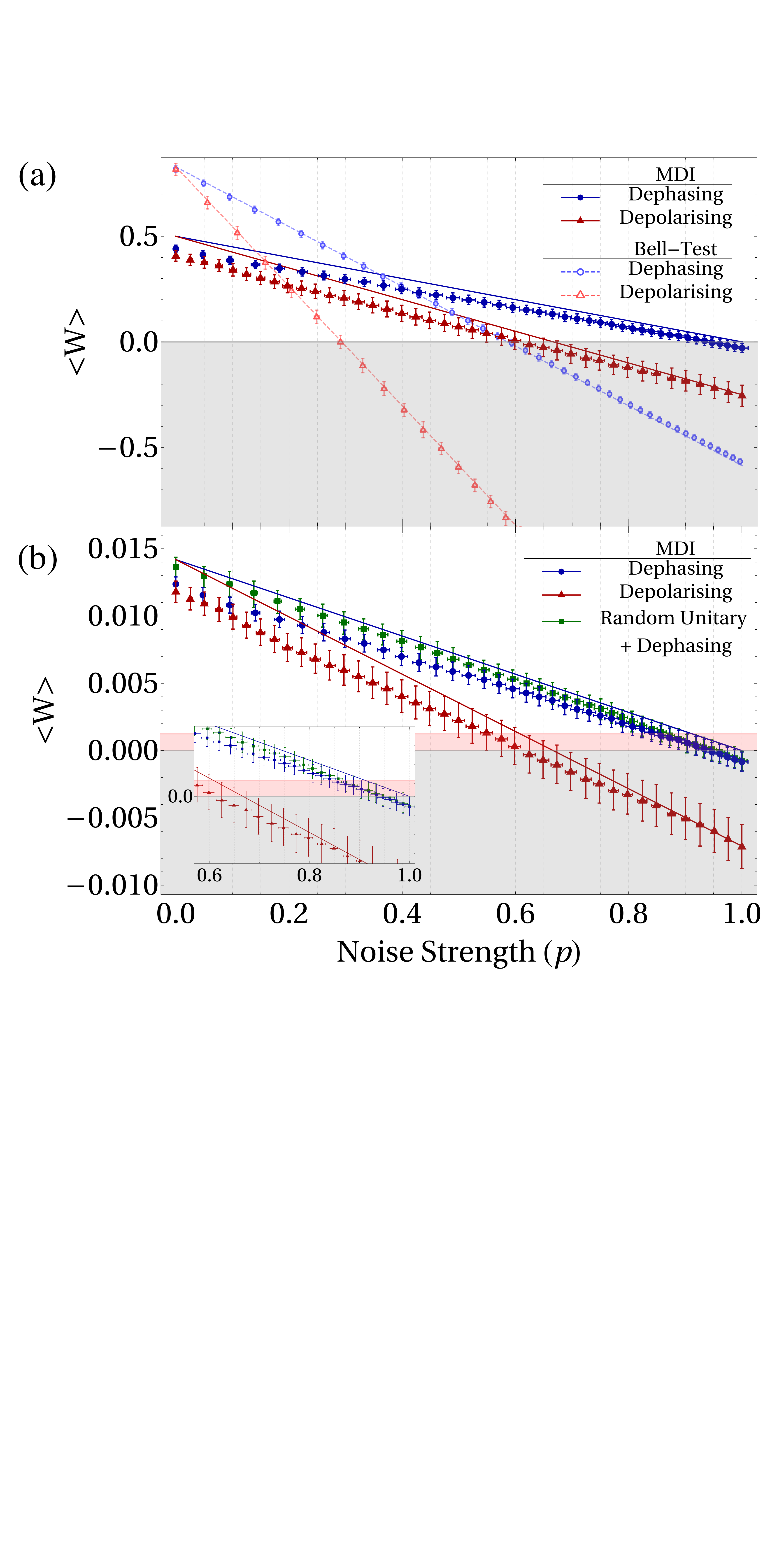}
  \end{center}
    \vspace{-2em}
    \caption{\textbf{Experimental results.}
   MDI entanglement witness measures (a) with and (b) without fair sampling assumption.
   Lines and points represent the theoretical prediction and the experimental data, respectively. The grey-shaded area corresponds to entanglement breaking channels, whilst the red-shaded area represents the actual threshold for entanglement breaking channels, taking into account Bob's optimal cheating strategy. 
   The discrepancy between theory prediction and data points is mainly due to imperfect two-photon interference on Bob's BS (we estimate a Bell-state measurement fidelity $\lesssim95\%$).
   Note that the theoretical lines are computed for illustration purposes only using the full knowledge of the setup, and therefore would not be available to Alice in the actual implementation of the protocol.
   The error bars in the data points represent $3\sigma$ statistical confidence regions obtained via Monte-Carlo re-sampling (n=$10^5$) assuming a Poissonian photon-counting distribution.}
    \label{fig:Results}
\end{figure}

Figure~\ref{fig:Results} (a) shows the result of our MDI channel verification for dephasing and depolarizing noise compared to a Bell-test approach with fair sampling assumption (i.e. neglecting the losses in the untrusted part of the setup). We show that, even in this idealized scenario, the MDI approach outperforms the Bell-test as it can certify quantumness for larger noise strength than the Bell-test (where the magnitude of noise each experimental approach can tolerate is computed from the intersection of the average payoff with the EB threshold).
In theory, the quantum nature of the depolarizing channel can be witnessed up to a noise level of $p<2/3$, while the Bell-based tests can only certify the channel up to $p\sim 0.29$. Surpisingly, the quantumness of a dephasing channel can be certified for any finite amount of noise, while a Bell-test approach can at best verify the channel up to $p \lesssim 0.58$.
Crucially, under ideal experimental conditions---i.e.\ no loss and perfect single photon sources---the best strategy Bob can use to convince Alice that he is in possession of a genuine quantum channel is to truthfully reveal the result of the joint measurement. Any other tactic would not maximize the payoff function~\cite{Rosset2017}.

\begin{figure}[t]
    \begin{center}
    \includegraphics[width=0.9\columnwidth]{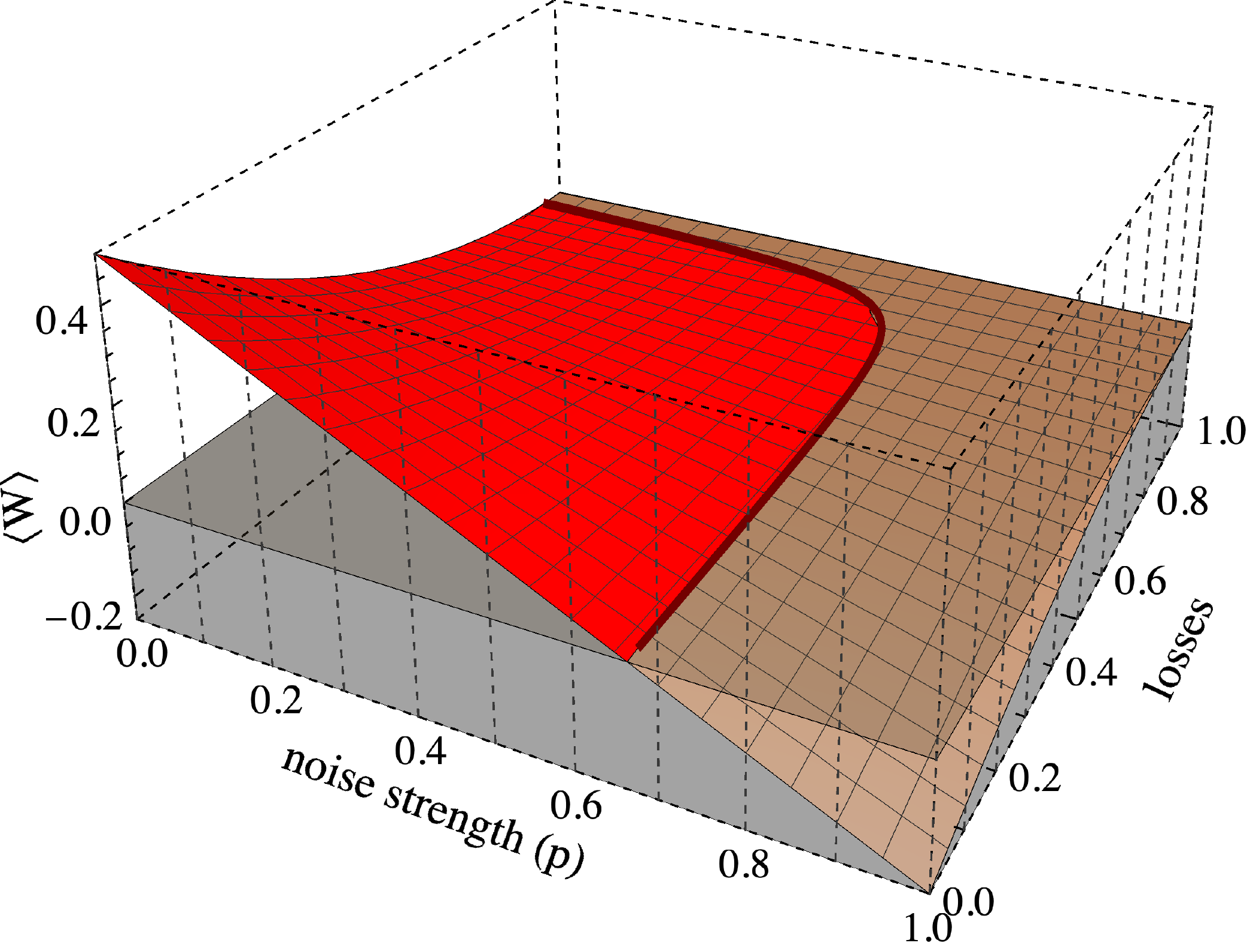}
    \end{center}
    \vspace{-2em}
    \caption{\textbf{MDI channel verification tradeoffs.} Shown is the ideal expectation value $\expec{W}$ of an MDI entanglement witness for an identity channel with depolarizing noise of strength $p$ and different amounts of loss. The multi-photon contribution is fixed at $\sim0.25\%$, as in our experimental setup, and results in a decreased size of the parameter region where certification of quantum behaviour is possible.}
    \label{fig:tradeoff}
\end{figure}

In order to guarantee device independence, the Bell-test approach would require very high detection efficiencies that are at best at the limit of current technical capabilities. 
The MDI approach, on the other hand, is less demanding in terms of experimental requirements.
The effects of losses and imperfections on our protocol are twofold. Firstly, lost photons lead to a decreased payoff, and secondly, Bob can exploit imperfections to cheat. Studying the latter possibility in some more detail, we note that most state-of-the-art photon sources suffer from a small probability of emitting multiple photons at a time, which Bob can exploit to extract information about the questions sent by Alice. Bob could then use this information to artificially inflate the payoff function by the following strategy: Whenever he gets no more than one photon in each question, he announces an unsuccessful projection on the singlet state (i.e.\ $b=1$). If he gets more than one photon in one of the questions and at least one in the other one, he can gather information on the question itself and perform a conveniently chosen local operation and classical communication (LOCC) positive-operator valued measure (POVM) to furtively inflate his payoff. In the Supplementary Materials, we derive the maximal payoff that Bob could achieve with a EB channel using these strategies and the known characteristics of Alice's photon source. By using this new threshold in our protocol we can reliably certify whether a channel is quantum, even if Bob is actively trying to cheat. Figure~\ref{fig:tradeoff} shows the theoretical trade-off between losses, noise and protocol success for depolarizing noise at a fixed ratio of multi-photon emission.

Taking into account both losses and multi-photon emissions according to our experimental parameters---overall heralding efficiency of $\sim17\%$ and multi-photon contribution of $\sim0.25\%$---puts us in a regime far beyond where a fully-DI approach would apply due to its sensitivity to loss. On the other hand, the MDI protocol reveals itself to be significantly more robust to such experimental imperfections, being still capable of certifying the nature of a quantum channel, as we show in Fig.~\ref{fig:Results}~(b).

So far, we have discussed MDI certification of a noisy identity channel. In practice, an imperfect channel might also apply some unknown unitary rotation to the stored qubit. In this case a nominally entanglement-preserving channel might appear to be EB due to the wrong choice of witness/payoff. In order to verify such channels, Alice uses a modified protocol, where she splits the answers obtained from Bob into two sets. The first set is used to reconstruct the channel's process matrix via quantum process tomography, and then computing the corresponding entanglement witness (as discussed, for example, in Ref.~\cite{GUHNE20091} or in the Supplemental Material~\ref{App:GamesFromData}). Alice can then perform the standard MDI verification with the adapted witness on the second data set. Since only the second stage of this extended protocol is MDI, Bob could attempt to cheat in the first stage. However, all this would achieve is that Alice computes a sub-optimal witness, which inevitably lowers the achievable payoff in the second stage. Bob's best strategy to have his channel certified is thus to broadcast the true outcome of the joint measurement he performs whilst Alice performs the process tomography of the channel, so that she can build the optimal witness for the channel at hand. 

Experimentally, the unknown unitary rotation was implemented by means of a HWP and a QWP at the output of the channel, and the MDI protocol was performed in the context of certifying its quantumness in the presence of dephasing noise. The results are shown in Fig.~\ref{fig:Results}(b) with losses and multi-photon corrections taken into account. We note that this protocol behaves as the standard MDI protocol for the identity channel, confirming the suitability of using a MDI approach in more complex scenarios where non-trivial noisy channels are involved.

% =========================================================
% Discussion
% =========================================================

\paragraph*{Discussion.---} 
We have shown a MDI framework that achieves quantum channel verification with sustained performance in the presence of noise and loss much beyond the capabilities of fully-DI methods, without the necessity of trusting the device under examination (as process-tomography-based methods would). With minimal demands on the trusted side (i.e. a single photon source), this method is ideally suited as a dependable benchmark for quantum memories and more general quantum channels. With the future vision of large scale quantum networks, this type of verification protocol can be a powerful tool for a skeptic who wishes to use the network, but does not necessarily trust the third party operating the network.
A natural extension of this work would be probing different properties of a quantum memory simultaneously. Fidelity, storage time and recall properties could be tested by changing the timing between the probe photons. On the theory side, the protocol could be extended to quantify the quantum nature of the channel instead of verifying it as has been done for the MDI quantification of entanglement~\cite{Shahandeh2017, Supic2017, Rosset2017b}; in this direction, one would need to use a capacity that quantifies specifically the quantum part of the channel: the negativity of the channel Choi state, or the quantum relative entropies~\cite{Yuan2019} could be used for that purpose.

\begin{acknowledgments}
We acknowledge that upon submission of this manuscript we became aware of similar work that has been recently carried out \cite{1906.09984}.
This work was supported by the UK Engineering and Physical Sciences Research Council (grant number EP/N002962/1). FG acknowledges studentship funding from EPSRC under grant no. EP/L015110/1. This project has received funding from the European Union’s Horizon 2020 research and innovation programme under the Marie Skłodowska‐Curie grant agreement No 801110 and the Austrian Federal Ministry of Education, Science and Research (BMBWF). It reflects only the author's view, the EU Agency is not responsible for any use that may be made of the information it contains.
Research at Perimeter Institute is supported by the Government of Canada through Industry Canada and by the Province of Ontario through the Ministry of Research and Innovation.
\end{acknowledgments}

\bibliography{MDIEW}
\bibliographystyle{apsrev4-1new}

\newpage
\clearpage
\newpage
\renewcommand{\theequation}{S\arabic{equation}}
\renewcommand{\thefigure}{S\arabic{figure}}
\renewcommand{\thetable}{S\arabic{table}}
\renewcommand{\thesection}{S\Roman{section}}
\setcounter{equation}{0}
\setcounter{figure}{0}
\setcounter{table}{0}
\setcounter{section}{0}

\section{Supplementary Material}

The Supplementary Material investigates the corrections required to the witness~\eqref{eq:payof}.
In Section~\ref{App:SQgames} we summarize the relevant definitions, describe for the first time methods to approximate the entanglement-breaking upper bound of a semi-quantum game, and how to construct optimal semi-quantum games from experimental data.
We compute the corrections for our particular witness in Section~\ref{App:CorrectionHigherOrder}, while explicit constructions for the cheating strategies are described in Section~\ref{App:CorrectionOurWitness}.
Finally, in Section~\ref{App:SparseWitness} we investiage the robustness of sparse witness decompositions to higher order events as well.

\subsection{Semi-quantum temporal games}
\label{App:SQgames}
We consider an arbitrary semi-quantum game; it is defined using sets of quantum input states $\{\xi_x\}_x$ and $\{\psi_y\}_y$, given as density matrices acting on Hilbert spaces $\mathcal{H}_\text{X}$ and $\mathcal{H}_\text{Y}$ respectively.
We write $P(b|x y)$ the probability distribution of getting the measurement outcome $b$ after the quantum inputs $\xi_x$ and $\psi_y$ have been entered into the box.
Later, we will consider the maximization of a given $\omega(b,x,y)$ payoff; currently, we want to characterize the correlations $P(b|x y)$ that Bob can achieve using various resources.
When Bob has access to a quantum channel $\mathcal{N}$, he can use channel to store the first input state $\xi_x$.
When he receives the second input state $\psi_y$, he retrieves the state $\mathcal{N}(\xi_x)$ from the channel and measures them jointly with the positive-operator valued measure (POVM) $\{B_b\}_b$, so that the probability of the outcome $b$ given input state indices $x$ and $y$ is:
\begin{equation}
  \label{Eq:SQUseChannel}
P(b|xy) = \mathrm{tr} \left [ B_b \left (\mathcal{N}(\xi_x) \otimes \psi_y \right) \right]\;.
\end{equation}
Note that Bob can pre-process the first input state before using the quantum channel $\mathcal{N}$; we refer the reader to~\cite{Rosset2017} for the full details.
In general, Bob's procedure can be seen as an effective joint measurement $\{\Pi_b\}_b$ acting on $\mathcal{H}_\text{X} \otimes \mathcal{H}_\text{Y}$, so that
\begin{equation}
  \label{Eq:Tomography}
P(b|xy) = \mathrm{tr} \left [ \Pi_b \left (\xi_x \otimes \psi_y \right ) \right ]\;.
\end{equation}
When the sets of input states $\{\xi_x\}_x$ and $\{\psi_y\}_y$ are tomographically complete, the correlations $P(b|xy)$ fully describe this effective POVM.

\subsubsection{Entanglement breaking channels}
We now consider the case when the quantum channel is entanglement breaking.
Such channels can be modeled as measure-and-prepare channels: instead of storing the input state $\xi_x$, the channel performs a quantum measurement on $\xi_x$ with outcome $\lambda = 1, \ldots, L$.
This classical outcome is stored in a classical memory, and a state $\xi'_x$ approximating $\xi_x$ is prepared at retrieval time.
In our framework, the state retrieved from the channel is measured jointly with the second input state $\psi_y$.
In the entanglement breaking case, the distributed POVM $\{\Pi_b\}_b$ is a 1-way local operation and classical communication (LOCC) POVM of the form
\begin{equation}
  \label{Eq:BobEBPOVM}
  \Pi_b = \sum_{\lambda} (X_{\lambda} \otimes Y_{b | \lambda})\, ,
\end{equation}
where $\{ X_{\lambda} \}_{\lambda}$ is a POVM acting on $\mathcal{H}_\text{X}$, and $\{ Y_{b | \lambda} \}_b$ is a POVM acting on $\mathcal{H}_\text{Y}$ for each $\lambda$:
\begin{equation}
  \sum_\lambda X_\lambda = 1, \qquad \sum_b Y_{b|\lambda} = 1,\;\forall \lambda \,.
\end{equation}
We write $\mathbf{\Pi}_\text{EB}$ the set of such distributed POVMs.
We also remark that the POVM elements of all entanglement-breaking $\{\Pi_b\}_b$ are separable.
This is how, in essence, MDIEWs can be constructed directly from a decomposition of entanglement witnesses over the quantum input states.
We now turn to computable relaxations of the set $\mathbf{\Pi}_\text{EB}$.

\subsubsection{Semidefinite relaxations for entanglement breaking semi-quantum devices}
\label{App:SemidefiniteRelaxations}

Inspired by~\cite{Doherty2004}, we construct semidefinite relaxations of the set $\mathbf{\Pi}_\text{EB}$ by observing that the form~\eqref{Eq:SQUseChannel} admits symmetric extensions.
In particular, we can define a POVM acting on $\mathcal{H}_\text{X} \otimes \mathcal{H}_{\text{Y}_1} \otimes \mathcal{H}_{\text{Y}_2}$, where both $\mathcal{H}_{\text{Y}_1}$ and $\mathcal{H}_{\text{Y}_2}$ are isomorphic to $\mathcal{H}_\text{Y}$:
\begin{equation}
  \label{symext}
  \overline{\Pi}_{b_1 b_2} = \sum_{\lambda} (X_{\lambda} \otimes Y_{b_1 | \lambda} \otimes Y_{b_2 | \lambda})\, ,
\end{equation}
and that POVM $\{\overline{\Pi}_{b_1 b_2}\}_{b_1 b_2}$ obeys the following conditions.
First, that POVM is symmetric.
Let $S$ be the operator that swaps the spaces $\mathcal{H}_{\text{Y}_1}$ and $\mathcal{H}_{\text{Y}_2}$. We have
\begin{equation}
  \label{conssym}
  S \overline{\Pi}_{b_1 b_2} S^\dagger = \overline{\Pi}_{b_2 b_1}, \quad \forall b_1, b_2\;.
\end{equation}
Second, the POVM elements and their partial transposes are all semidefinite positive:
\begin{equation}
  \label{consppt}
  \overline{\Pi}_{b_1 b_2} \ge 0, \quad \overline{\Pi}_{b_1 b_2}^{\top_\text{X}} \ge 0, \quad \overline{\Pi}_{b_1 b_2}^{\top_{\text{Y}_1}} \ge 0
\end{equation}
(the other partial transposes are redundant by symmetry).
Finally, the extended POVM $\{ \overline{\Pi}_{b_1 b_2} \}$ obeys nonsignaling constraints: from~Eq.~\eqref{symext}, we see that after summing over $b_2$, the POVM
\begin{equation}
  \label{consns}
  \sum_{b_2} \overline{\Pi}_{b_1 b_2} = \sum_{\lambda} (X_{\lambda} \otimes Y_{b_1 | \lambda} \otimes 1) = \Pi_{b_1} \otimes \mathbbm{1}
\end{equation}
cannot possibly depend on the input given in $\mathcal{H}_{\text{Y}_2}$.
Finally, we need to enforce normalization of $\overline{\Pi}_{b_1 b_2}$ by requiring $\sum_{b_1 b_2} \overline{\Pi}_{b_1 b_2} = \mathbbm{1}$; that constraint can be replaced by $\sum_{b_1 b_2} \overline{\Pi}_{b_1 b_2} = \alpha \mathbbm{1}$ with $\alpha \ge 0$ when the conic span is required.

More refined relaxations are easily obtained by using additional copies of the second subsystem.

\subsubsection{Semi-quantum games and maximal payoffs}
A semi-quantum game is described by the payoff coefficients $\omega(b,x,y)$, so that the expected payoff is
\begin{equation}
  \label{expectedpayoff}
  W = \sum_{bxy} \omega(b,x,y) P(b | x y)\,.
\end{equation}

To any semiquantum game, we can associate the following maximal payoffs.
First, $W_\text{EB}$ is the maximal payoff that can be obtained when the channel Bob uses is entanglement breaking as in~\eqref{Eq:BobEBPOVM}.
We also write $W_\text{Q}$ the maximal payoff that can be achieved with a quantum channel as in~\eqref{Eq:SQUseChannel} or its generalizations.
Finally, we consider the situation where Bob has full knowledge of the indices $x$ and $y$, instead of only having access to the states $\xi_x$ and $\psi_y$.
In that case, Bob can simply store the classical label $x$ in a classical memory, and process it along with $y$ to output $b^\star(x,y) = \myop{argmax}_b \omega(b,x,y)$, thus achieving the maximal algebraic value $W_\mathrm{ALG}$ of the payoff:
\begin{equation}
  W_{\mathrm{ALG}} = \sum_{x y} \max_b \omega(b,x,y) \;.
\end{equation}

In the main text, we discussed a construction~\cite{Rosset2017} of payoffs from entanglement witnesses, in which case we have $W_\text{EB} = 0$ and $W_\text{Q} > 0$ by construction.
Proving convergence of the sequence of such relaxations is left open.

\subsubsection{Approximating $W_\text{EB}$}
\label{Sec:ApproximatingEB}
A crude upper bound on $W_\text{EB}$ is given by the algebraic bound $W_\text{ALG}$.
Otherwise, the semidefinite relaxations described in Section~\ref{App:SemidefiniteRelaxations} enable the computation of finer upper bounds on $W_\text{EB}$.
Indeed, one can simply maximize
\begin{equation}
  W_\text{EB}^\text{ub} = \max_{\{\Pi_b\}_b \in \tilde{\mathbf{\Pi}}_\text{EB}} \sum_{bxy} \omega(b,x,y) \mathrm{tr}\left[\Pi_b \left(\xi_x \otimes \psi_y\right)\right]
\end{equation}
where $\tilde{\mathbf{\Pi}}_\text{EB}$ is a semidefinite representable set coming from a symmetric extension of some order.
This upper bound is then compared with the lower bound obtained by the following search for explicit realizations.
We fix the maximal cardinality $L$ of the classical information, and optimize over POVMs $\{X_\lambda\}_\lambda$ and $\{Y_{b|\lambda}\}_b$.
The resulting optimization problem is not convex; nevertheless, by alternating optimization on $\{X_\lambda\}_\lambda$ and $\{Y_{b|\lambda}\}_b$, we can reduce the problem to a tractable optimization.
For that purpose, we start by sampling a random POVM $\{X^0_\lambda\}_\lambda$, and solve the following convex optimization problem:
\begin{align}
  W_\text{EB}^\text{lb} = \max_{\{Y_{b|\lambda}\}} & \sum_{\lambda bxy} \omega(b,x,y) ~ \mathrm{tr}[X^0_\lambda ~ \xi_x] ~ \mathrm{tr}[Y_{b|\lambda} ~ \psi_y] \\
  \text{subject to } & \sum_b Y_{b|\lambda} = 1, \forall \lambda \\
                           & Y_{b|\lambda} \ge 0, \forall b, \lambda\;.
\end{align}
Now, fixing the set of POVMs $\{Y_{b|\lambda}\}_b$, we solve the convex optimization problem:
\begin{align}
  W_\text{EB}^\text{lb} = \max_{\{X_\lambda\}} & \sum_{\lambda bxy} \omega(b,x,y) ~ \mathrm{tr}[X_\lambda ~ \xi_x] ~ \mathrm{tr}[Y_{b|\lambda} ~ \psi_y] \\
  \text{subject to } & \sum_\lambda X_{\lambda} = 1, \\
                           & X_\lambda \ge 0, \forall  \lambda\;.                        
\end{align}
We alternate both optimizations until no improvement in the objective value is observed. At any point in this iterative process, the variables $\{X_\lambda\}$ and $\{Y_{b|\lambda}\}$ represent a valid explicit model for semiquantum correlations out of an entanglement breaking channel, and thus $W_\text{EB}^\text{lb} \le W_\text{EB}$. The whole optimization procedure should be repeated a few times with different initial random POVMs to avoid getting trapped in a non-optimal local maximum (which happens in the majority of tries).

\subsubsection{Constructing semi-quantum games from experimental data}
\label{App:GamesFromData}
Let assume we estimate correlations $P(b|xy)$ from experimental data.
We sketch below a translation of the method of~\cite{Verbanis2015, Rosset2017b} to the case of temporal correlations.

First, one has to ensure that there exists a POVM with positive semidefinite elements $\{\Pi_b\}_b$ satisfying~\eqref{Eq:Tomography}; a possible regularization procedure is described in~\cite{Rosset2017b}.
Then, one solves the following feasibility problem:
\begin{align}
  \text{minimize } & 0 \nonumber \\
  \text{over } & \{\Pi_b\}_b \in \hat{\mathbf{\Pi}}_\text{EB} \nonumber \\
  & \mathrm{tr} \left[ \Pi_b \left(\xi_x \otimes \psi_y \right) \right] = P(b|xy)
\end{align}
where $\hat{\mathbf{\Pi}}_\text{EB}$ is the conic span of one of the semidefinite relaxations of $\mathbf{\Pi}_\text{EB}$ described in Section~\ref{App:SemidefiniteRelaxations}.
The dual of this problem maximizes the objective $d = \sum_{bxy} \omega(b,x,y) P(b|xy)$; by weak duality, we have $d \le 0$ when the primal problem is feasible, and thus $d > 0$ is a signature of $P(b|xy)$ describing correlations coming from a non-entanglement-breaking channel.
Then the coefficients $\omega(b,x,y)$ form the payoff of a semi-quantum game (for details about weak duality, see the Appendix of~\cite{Rosset2017b}).

\subsection{Correction for higher-order emissions}
\label{App:CorrectionHigherOrder}
In our implementation, there is a large probability that no input state is produced although a heralding event is recorded.
There is a small probability of obtained a single photon, and a very small probability of producing multiple copies.
We write $n_1, n_2$ the number of copies of the first and second input state respectively.
By considering the experimental parameters of the source (PDC squeezing parameter equal to $4.4\cdot 10^{-3}$, ``trusted'' heralding efficiency in Alice's part of the setup of $\sim35\%$, $80$ MHz laser repetition rate and detector dead time of $\sim50$ ns), we can deduce the probability of $(n_1, n_2)$ heralded copies produced from the experimental device and successfully sent to Bob, which are presented in Table~\ref{Tab:HigherOrder}.
We compute the maximal quantum payoff achievable; note that our experimental implementation only performs the optimal measurement for $n_1 = n_2 = 1$, whereas higher order events will add noise to the result.
As we want to falsify the case where Bob has only an entanglement-breaking channel at his disposal, we compute the maximal entanglement-breaking payoff $W_{\text{EB}}$ when using the multiple copies available to cheat:
\begin{equation}
\begin{split}
W_{\text{EB}}=\sum_{(n_1,n_2)}  P(n_1,n_2) W_{\text{EB}}^{(n_1,n_2)} \, ,
\end{split}
\end{equation}
where an explicit construction of those cheating strategies are given in the next section (\ref{App:CorrectionOurWitness}).
The non-EB nature of the channel is therefore verified whenever 
\begin{equation}
\begin{split}
\frac{W}{\eta_{\text{trusted}}^2} > W_{\text{EB}} \, ,
\end{split}
\end{equation}
where the left hand side is the measured witness renormalized with the trusted heralding efficiency per question $\eta_{\text{trusted}}$. In our experimental implementation the EB boundary is $W_{\text{EB}}\approx 1.25\cdot10^{-3}$.

\begin{table}[]
\begin{tabular}{|l|l|l|l|l|}
\hline
$(n_1,n_2)$ & $W_\text{Q}$ & $W_\text{EB}^{(n_1,n_2)}$ & $P (n_1, n_2)$ \\ \hhline{|=|=|=|=|}
\shortstack{$(0,0)$ \\ $(n_1>0,0)$ \\ $(0,n_2>0)$} & $0$ & $0$ & $0.75$ \\ \hline
$(1,1)$ & $1/2$ & $0$ & $0.24$ \\ \hline
$(1,2)$ & $2/3$ & $1/3$ & $1.5 \cdot 10^{-3}$ \\ \hline
$(1,3)$ & $\cong 0.947$ & \shortstack{ $\in[\sqrt{2}/3, 0.474]$ \\ $\in[0.471, 0.474]$ }  & $5 \cdot 10^{-6}$ \\ \hline
$(2,1)$ & $2/3$ & $1/2$ & $1.5 \cdot 10^{-3}$ \\ \hline
$(2,2)$ & $10/9$ & $2/3$ & $8 \cdot 10^{-6}$  \\ \hline
$(3,1)$ & $\cong 0.947$ & \shortstack{ $2\sqrt{2}/3$ \\ $\cong 0.943$} & $5 \cdot 10^{-6}$  \\ \hline
other & $\le 3/2$ &  $\le 3/2$ & $9 \cdot 10^{-8}$ \\ \hline
\end{tabular}
\caption{
  \label{Tab:HigherOrder}
  The maximal payoff $W_\text{Q}$ can be obtained with the ideal identity channel that fully preserves quantum states.
  We present the optimal payoff $W_\text{EB}$ obtained by using an entanglement breaking channel.
  When the lower and upper bounds match within machine precision ($\epsilon = 10^{-7}$), we present a single number, otherwise the lower/upper bound interval. $P (n_1, n_2)$ are the probabilities that Bob receives $(n_1,n_2)$ copies of the questions at each run of the protocol.
}
\label{tab:Web}
\end{table}

\subsection{Cheating strategies for the game implemented}
\label{App:CorrectionOurWitness}
We now consider the semi-quantum game described in the main text with payoff~\eqref{eq:payof}.
Let us assume that instead of a single copy of the input states $\xi_x$ and $\psi_y$, multiple copies of those states are prepared and sent to the untrusted devices:
\begin{equation}
  \tilde{\xi}_x = \xi_x^{\otimes n_1}, \qquad \tilde{\psi}_y = \psi_y^{\otimes n_2}
\end{equation}
where $n_1, n_2 = 1, 2, 3, \ldots$ denote the number of copies, which happens with probability $P (n_1, n_2)$.
Bob will learn more information about the input when several copies are provided. We are interested in the maximal payoff possibly obtained when the channel is entanglement breaking.
We use the two methods described above to obtain lower and upper bounds on $W_\text{EB}$ for the payoff~\eqref{eq:payof}.

In Table~\ref{tab:Web2} we compute the entanglement-breaking bound $W_{\mathrm{EB}}^{(n_1, n_2)}$ for
$n_1 + n_2 \le 4$ copies of $\xi$ and $\psi$ respectively.
We also compute the algebraic bound $W_\mathrm{ALG}$, which will be used as an upper bound for $W_\mathrm{EB}^{(n_1,n_2)}$ for $n_1 + n_2 > 4$.

\begin{table}[H]
  \centering
\begin{tabular}{ccccccc}
\hline\hline
 $W_{\mathrm{EB}}^{(\leqslant 1)}$ & $W_{\mathrm{EB}}^{(1,2)}$ & $W_{\mathrm{EB}}^{(1,3)}$ & $W_{\mathrm{EB}}^{(2,1)}$ & $W_{\mathrm{EB}}^{(2,2)}$ & $W_{\mathrm{EB}}^{(3,1)}$ & $W_{\mathrm{ALG}}$\\
  \hline
  0 & $1/3$ & $0.474^*$ & $1/2$ & $2/3$ & $\frac{2}{3}\sqrt{2} \cong 0.943$ & $3/2$ \\
  \hline\hline
\end{tabular}
\caption{$W_{\mathrm{EB}}^{(\leqslant 1)}$ contains the cases $(n_1, n_2) \in \{ (0, 0), (0, 1), (1, 0), (1, 1) \}$. Except for the case marked $^*$, the lower and upper bounds match within a precision of $\epsilon = 10^{-7}$.}
\label{tab:Web2}
\end{table}

In the next subsection, we will compute explicitly the optimal strategies achieving $W_\mathrm{EB}^{(n_1, n_2)}$ and show that Bob's strategy depends on the number of incoming photons $n_1$; thus optimal cheating requires a non demolishing photon counting measurement. In our

\subsection{Explicit strategies for higher-order emissions}

We now describe the optimal cheating strategy for various numbers of copies $(n_1, n_2)$.
In this part of the manuscript, we work with state vectors instead of density matrices and denote the quantum input states by $| \xi_x \rangle \in \mathbbm{C}^2$.
We fix the basis in which the four non-orthogonal inputs are expressed.
With $\textbf{i} = \sqrt{- 1}$, $\alpha = 1 / 2 + 1 / \sqrt{12}$, we write in the computational basis:
\begin{equation}
  | \xi_0 \rangle = \sqrt{\alpha} | 0 \rangle + \sqrt{1 - \alpha} e^{\textbf{i} \pi / 4} | 1 \rangle,
\end{equation}
and the other states are given by $| \xi_i \rangle = \sigma_i | \xi_0 \rangle \text{}$ where $\sigma_1$, $\sigma_2$ and $\sigma_3$ are the Pauli X,Y and Z matrices respectively.
Note that the states $| \phi_y \rangle$ are taken from the same set, and thus we reuse $| \phi_y \rangle = | \xi_y \rangle$ for simplicity.
At the single copy level, the global phases in those definitions do not matter; however, such phases will affect the derivations below.
For a number of copies $n$ we define $| \xi^{n, 0}_i \rangle = | \xi_i \rangle^{\otimes n}$.
\subsubsection{Method}

For each pair $(n_1, n_2)$ under investigation, we worked as follows.
We computed a best effort locally optimal explicit strategy using the method described in Section~\ref{Sec:ApproximatingEB}, by using the iterative procedure on random starting points, with various cardinalities for $\lambda$.
In all cases considered, the optimal cardinality for $\lambda$ is $\lambda \in \{ 0, 1, 2, 3 \}$, hinting at a close relation between $\lambda$ and the input state $x$.
Indeed, we observe that the POVM elements $\{ X_{\lambda} \}_{\lambda}$ has the same symmetries as the set $\{ | \xi_i^{n, 0} \rangle \}_i$.
This led us to the following Ansatz, the optimality of which was verified using the numerical upper bound obtained by semidefinite relaxations.

Let $S_n$ be the symmetric group of degree $n$, and $\pi : S_n \rightarrow(\mathbbm{C}^2)^{\otimes n}$, $g \mapsto \pi_g$ the representation that permutes subsystems.
Then
\begin{equation}
  \label{Eq:Relations1} \forall g \in S_n, \quad \pi_g | \xi_i^{n, 0} \rangle = | \xi_i^{n, 0} \rangle,
\end{equation}
and
\begin{equation}
  \label{Eq:Relations2} \sigma_i | \xi_0^{n, 0} \rangle = | \xi_i^{n, 0} \rangle, \quad i = 1, 2, 3.
\end{equation}
We define
\begin{equation}
  \left| \overline{\xi}_0 \right\rangle = \sqrt{1 - \alpha} e^{- \textbf{i} \pi / 4} | 0 \rangle - \sqrt{\alpha} | 1 \rangle,
\end{equation}
with $\left| \overline{\xi}_i \right\rangle = \sigma_i \left| \overline{\xi}_0 \right\rangle$ so that $\left| \overline{\xi}_i \right\rangle$ is orthogonal
to $| \xi_i \rangle$:
\begin{equation}
  \left| \left\langle \overline{\xi}_i \middle| \xi_j \right\rangle \right|^2 = \frac{2}{3} (1 - \delta_{i j}) .
\end{equation}
On the Bloch sphere, the $| \xi_i \rangle$ form a tetrahedron, and $\left| \overline{\xi}_i \right\rangle$ its reflection.
We note that $\left| \overline{\xi}_i \right\rangle^{\otimes n}$ obeys the invariance requirements~(\ref{Eq:Relations1})-(\ref{Eq:Relations2}) by construction.
In general, for $n = m + \overline{m}$, the set of states $\left\{ \xi_i^{m, \overline{m}} \right\}_i$
\begin{equation}
  \left| \xi_i^{m, \overline{m}} \right\rangle = \frac{1}{n!} \sum_{g \in S_n} \pi_g \left( | \xi_i \rangle^{\otimes m} \left| \overline{\xi}_i \right\rangle^{\otimes \overline{m}} \right)
\end{equation}
fit those requirements as well.
We now consider the cases $n_1 = 1, 2, 3$ separately, building our POVM elements from the states $\left\{ \left| \xi_i^{m, \overline{m}} \right\rangle \right\}_i$.
The part of our POVMs that acts on the antisymmetric subspace is irrelevant, and we chose to distribute it uniformly among elements; we define it formally as:
\begin{equation}
  \label{Eq:AntiSymProj}
  \Pi^n_\mathrm{anti} = \mathbbm{1} - \sum_{m + \overline{m} = n} \left| \xi^{m, \overline{m}}_0 \right\rangle \left\langle \xi_0^{m, \overline{m}}\right|
\end{equation}

\subsubsection{Strategies for $n_1 = 1$}

For the case $n_1 = 1$ in our Ansatz, we consider the first POVM $\{X_{\lambda} \}_{\lambda}$.
In the cases $n_2 = 1, 2, 3$ considered, that optimal POVM was always build out of $\left| \overline{\xi}_i \right\rangle = | \xi_i^{0, 1} \rangle$
\begin{equation}
  X^1_{\lambda} = \left| \overline{\xi}_{\lambda} \right\rangle \left\langle
  \overline{\xi}_{\lambda} \right| / 2
\end{equation}
where the POVM element $X_{\lambda}^1$ has zero overlap with $| \xi_{\lambda} \rangle$, and thus the outcome $\lambda$ communicates that the input state is
{\em not} $| \xi_{\lambda} \rangle$.
What is being done with this information for the second measurement, described by a set of POVMs $\{Y_{b| \lambda} \}_b$ indexed by $\lambda$ with outcome $b$?
It would make sense to maximize the overlap of $Y_{0 | \lambda}$ with the second input state $| \xi_{\lambda}^{n_2, 0} \rangle$, with $Y_{1 | \lambda} = \mathbbm{1} - Y_{0 | \lambda}$.

For $n_2 = 1, 2$, this is exactly what happens.
The optimal POVM elements are given by
\begin{equation}
  Y^{(1, 1)}_{0 | \lambda} = | \xi^{1, 0}_{\lambda} \rangle \langle \xi^{1, 0}_{\lambda} |, \quad Y^{(1, 2)}_{0 | \lambda} = | \xi_{\lambda}^{2, 0} \rangle \langle \xi^{2, 0}_{\lambda} | .
\end{equation}
We then get $W_{\text{EB}}^{(1, 1)} = 0$ and $W_{\text{EB}}^{(1, 2)} = 1 / 3$, which is verified to be optimal up to solver precision using a symmetric extension of level 3.

For $n_2 = 3$, using the same strategy $(Y_{0 | \lambda}^{(1, 3)} = | \xi_{\lambda}^{3, 0} \rangle \langle \xi_{\lambda}^{3, 0} |)$ gives a payoff $W^{(1, 3)} = 4 / 9 \cong 0.4444$ which is suboptimal.
Indeed, keeping $\{X_{\lambda}^1 \}_{\lambda}$ fixed as above, the corresponding optimal second measurement is obtained by semidefinite programming, and corresponds to
\begin{equation}
  Y_{0 | \lambda}^{(1, 3)} = | \gamma_{\lambda} \rangle \langle \gamma_{\lambda} |,
\end{equation}
with\footnote{Note that the relative phase between $| \xi^{3, 0} \rangle$ and
$| \xi^{0, 3} \rangle$ depends on the global phase used in the definition of
the $\left| \overline{\xi}_i \right\rangle$.}
\begin{equation}
  \label{Eq:gamma} | \gamma_{\lambda} \rangle = \beta | \xi^{3, 0} \rangle + \sqrt{1 - \beta} e^{- \textbf{i} \pi / 4} | \xi^{0, 3} \rangle\;,
\end{equation}
$beta = 1/\sqrt{3} + 1/\sqrt{6}$, which provides $W^{(1, 3)} = \sqrt{2} / 3 \cong 0.4714$.
A gap still remains with the upper bound obtained through symmetric extensions of level 3:
\begin{equation}
  \frac{\sqrt{2}}{3} \leqslant W_{\text{EB}}^{(1, 3)} \lesssim 0.47402.
\end{equation}
As no better explicit strategies were found after extensive search, we believe the left bound to be tight.
However, we use the upper bound from symmetric extensions in our analysis to be safe.

\subsubsection{Strategies for $n_1 = 2$}
For $n_1 = 2$ with $n_2 = 1, 2$, the optimal first measurement becomes
\begin{equation}
  X_{\lambda}^2 = \frac{3 | \xi_{\lambda}^{2, 0} \rangle \langle \xi_{\lambda}^{2, 0} | + \Pi_\mathrm{anti}^2}{4},
\end{equation}
where the measurement outcome $\lambda$ indicates that the input state is now {\em likely} to be $| \xi^2_{\lambda} \rangle$ and $\Pi^2_\mathrm{anti}$ is the projector on the antisymmetric subspace of Eq.~\eqref{Eq:AntiSymProj}.
When $n_1 = 1$, it seems to make sense to minimize the overlap with $| \xi_{\lambda} \rangle$, indeed we observe that the optimal second measurement is
\begin{equation}
  \label{Eq:Y21} Y_{0 | \lambda}^{(2, 1)} = | \xi^{0, 1}_{\lambda} \rangle \langle \xi^{0, 1}_{\lambda} |
\end{equation}
providing $W_{\text{EB}}^{(2, 1)} = 1 / 2$.
For $n_2 = 2$, the optimal second measurement is actually of rank two, and written
\begin{equation}
  Y_{0 | \lambda}^{(2, 2)} = | \xi_{\lambda}^{0, 2} \rangle \langle   \xi_{\lambda}^{0, 2} | + | \xi_{\lambda}^{(1, 1)} \rangle \langle   \xi_{\lambda}^{(1, 1)} |,
\end{equation}
which gives $W_{\text{EB}}^{(2, 2)} = 2 / 3$.
Those bounds were verified up to machine precision using symmetric extensions of level 3.

\subsubsection*{Strategy for $n_1 = 3$ and $n_2 = 1$}

For $(n_1, n_2) = (3, 1)$, the optimal first measurement is projective
\begin{equation}
  X_{\lambda}^3 = | \gamma_{\lambda} \rangle \langle \gamma_{\lambda} | + \frac{\Pi_{\mathrm{anti}}^3}{4},
\end{equation}
using~(\ref{Eq:gamma}); while an outcome $i$ indicates that the input state is likely to be $| \xi^3_{\lambda} \rangle$, the state $| \gamma_{\lambda}
\rangle$ is not the one with maximal overlap with the $| \xi_{\lambda}^3 \rangle$.

As in~(\ref{Eq:Y21}), the second measurement minimizes the overlap with $|\xi_{\lambda} \rangle$:
\begin{equation}
  Y_{0 | \lambda}^{(3, 1)} = \left| \overline{\xi}^1_{\lambda} \right\rangle \left\langle \overline{\xi}^1_{\lambda} \right|
\end{equation}
which gives $W_{\text{EB}}^{(3, 1)} = 2 \sqrt{2} / 3 \cong 0.9428$.
This bound matches the upper bound obtained using symmetric extensions of level 3 up to machine precision.

\subsection{Sparse witness decomposition}
\label{App:SparseWitness}
\begin{figure}[htbp]
    \begin{center}
    \includegraphics[width=0.96\columnwidth]{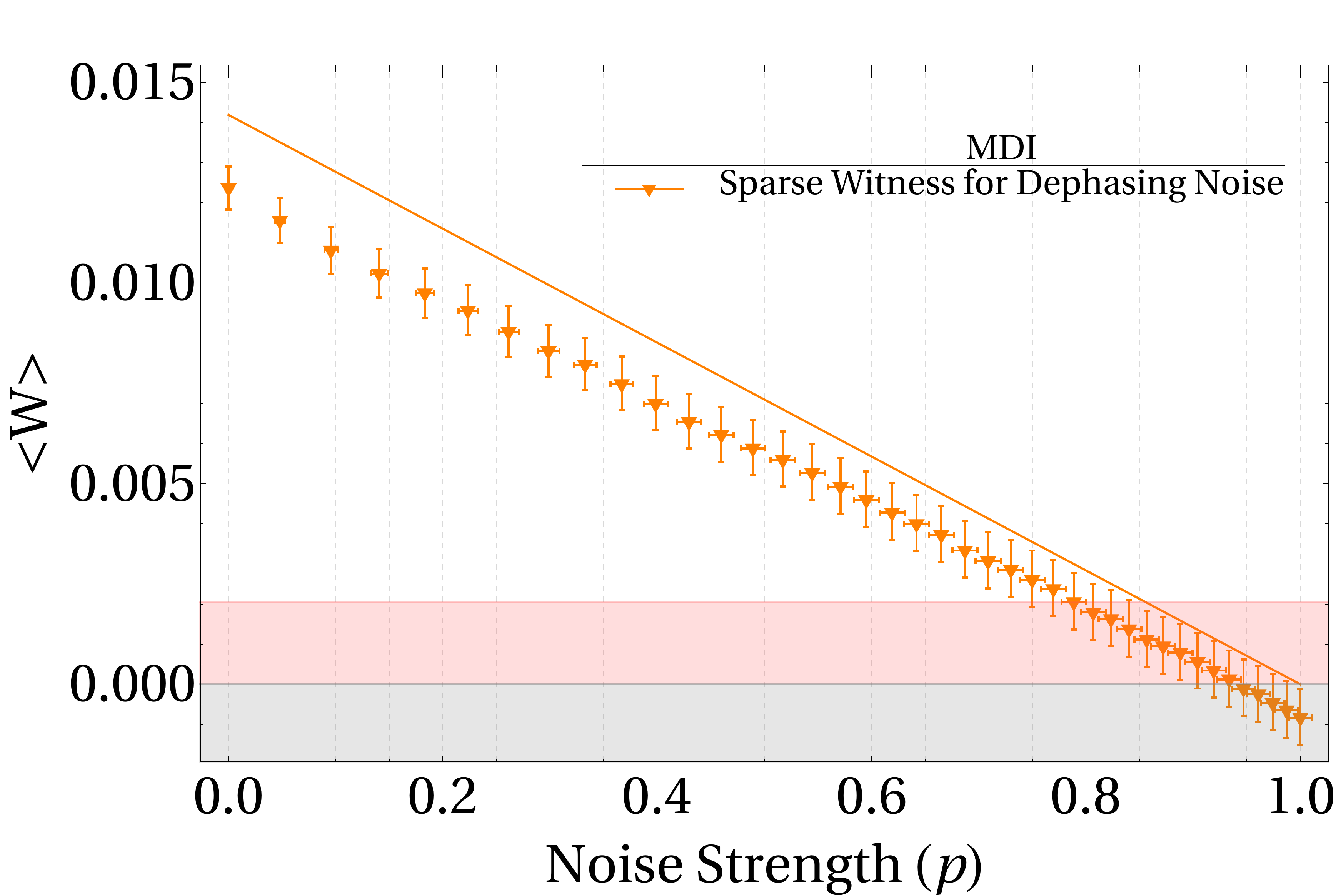}
    \end{center}
    \vspace{-2em}
    \caption{\textbf{Sparse decomposition implementation of the MDI protocol.}
    MDI entanglement witness without fair sampling assumption.
   Lines and points represent the theoretical prediction and the experimental data, respectively, and the shaded areas correspond to entanglement breaking channels. The red-shaded area takes into account cheating strategies. 
   The error bars in the data points represent $3\sigma$ statistical confidence regions obtained via Monte-Carlo re-sampling (n=$10^5$) assuming a Poissonian photon-counting distribution.}
    \label{fig:Sparse}
\end{figure}

The entanglement witness decomposition into the SIC states discussed in the main text is not unique: indeed, other decompositions into set of non-orthogonal states can be used. A relevant example is the so-called sparse decomposition, which minimises the number of questions needed for implementing the MDIEW protocol~\cite{Rosset2017}.
The sparse decomposition involves the  following set of quantum states for both the first and the second question:
\begin{equation}
    \begin{split}
    &\xi_0 = \frac{\id}{2}     \\
    &\xi_1 = \frac{\id + \sigma_x}{2}     \\
    &\xi_2 = \frac{\id + \sigma_y}{2}     \\
    &\xi_3 = \frac{\id + \sigma_z}{2} \\
    &\xi_4 = \frac{\id + (\sigma_x+\sigma_y+\sigma_z)/\sqrt{3}}{2} \, .
    \end{split}
    \label{eq:sparse}
\end{equation}
This set results in only six non-zero coefficients
\begin{align}
  \omega(0,0,0) & = 2(\sqrt{3} - 1), \nonumber \\
  \omega(0,0,4) = \omega(0,4,0) &= -\sqrt{3}, \nonumber \\
  \omega(0,1,1) = -\omega(0,2,2) = \omega(0,1,1) &= 1,
\end{align}
corresponding to six input pairs of questions that need to be measured.
We implemented the protocol in the presence of simulated dephasing noise, obtaining the results shown in Fig.~\ref{fig:Sparse}.
As expected, the entanglement witness behaves qualitatively analogously to the SIC decomposition chosen in main text.
  We note that, despite the reduced number of joint measurements required by the sparse decomposition, the set of states in \eqref{eq:sparse} can't be easily generated with our parametric source, while requires the generation of different pure states that are therefore statistically mixed in post-selection for obtaining the needed set.
  While this is theoretically equivalent to directly generate the correct states, it's not a convenient strategy in general when considering our photon source, and we therefore showed the decomposition into the SIC set as our main result.
  While the sparse decomposition as the same maximal quantum payoff $W_\text{Q}^\text{sparse} = 1/2$, the resistance to cheating strategies is worse, see Table~\ref{Tab:HigherOrderSparse}.
\begin{table}[]
\begin{tabular}{|l|l|l|}
\hline
$(n_1,n_2)$ & $W_\text{EB}$ & $W_\text{EB}^\text{sparse}$ \\ \hhline{|=|=|=|}
  \shortstack{$(0,0)$ \\ $(0,1)$ \\ $(1,0)$}
            & $0$ & $0$ \\ \hline
$(1,1)$ & $0$ & $0$ \\ \hline
$(1,2)$ & $1/3$ & $\cong 0.757$ \\ \hline
$(1,3)$ & $\in [0.471, 0.474]$ & $\in[1.113, 1.119]$ \\ \hline
$(2,1)$ & $1/2$ & $\in[0.491, 0.607]$ \\ \hline
$(2,2)$ & $2/3$ & $\cong 1.334$ \\ \hline
$(3,1)$ & $\cong 0.943$ & $\in [1.068, 1.075]$ \\ \hline
$n_1+n_2>4$ &  $\le 3/2$ & $\le \sqrt{12} \cong 3.46$ \\ \hline
\end{tabular}
\caption{
  \label{Tab:HigherOrderSparse}
  Resistance to higher order events for the sparse witness decomposition, as evidenced by the entanglement breaking maximal payoff $W_\text{EB}^\text{sparse}$, compared to bound for the standard decomposition $W_\text{EB}$ computed in Table~\ref{Tab:HigherOrder}.
}
\end{table}

\end{document}